\documentclass[12pt]{article}
\usepackage{amsmath,amssymb}
\usepackage{graphicx}
\allowdisplaybreaks[1]

\makeatletter
 
  \@addtoreset{equation}{section}
 \makeatother
\newcommand{\bfa}{{\mathbf a}}

\newcommand{\bfk}{{\mathbf k}}

\newcommand{\bfm}{{\mathbf m}}
\newcommand{\bfn}{{\mathbf n}}

\newcommand{\bfr}{{\mathbf r}}

\newcommand{\bz}{\bar{z}}

\newcommand{\bphi}{{\bar{\phi}}}

\newcommand{\tn}{\tilde{n}}

\newcommand{\cN}{{\cal N}}
\newcommand{\cO}{{\cal O}}

\newcommand{\vk}{\vec{k}}

\newcommand{\av}{\mbox{\boldmath$a$}}

\newcommand{\Z}{\mathbb{Z}}

\newcommand{\N}{\mathbb{N}}

\newcommand{\nn}{\nonumber}

\newcommand{\Tr}{{\rm Tr}\,}

\newcommand{\e}{\epsilon}

\begin{document}

\begin{titlepage}

\setcounter{page}{0}
\renewcommand{\thefootnote}{\fnsymbol{footnote}}

\vspace{20mm}

\begin{center}
{\large\bf
Instanton Counting and Dielectric Branes
}

\vspace{20mm}
{
So Matsuura\footnote{{\tt matsuura@nbi.dk}}}\\
\vspace{10mm}

{\em The Niels Bohr International Academy, 
The Niels Bohr Institute, \\
Blegdamsvej 17, DK-2100 Copenhagen, Denmark}

\end{center}

\vspace{20mm}
\centerline{{\bf Abstract}}
\vspace{10mm}
We consider the Hanany-Witten type brane configuration 
in a background of RR 4-form field strength 
and examine the behavior of 
Euclidean D0-branes propagating between two NS5-branes. 
We evaluate the partition function of the D0-branes 
and show that it coincides with the Nekrasov partition 
function of instantons for four-dimensional $\cN=2$ 
supersymmetric Yang-Mills theory. 
In this analysis, the Myers effect plays a crucial role. 
We apply the same method to the brane configuration 
realizing four-dimensional $\cN=2$ theory with hypermultiplets 
in the fundamental representation and reproduce 
the corresponding Nekrasov partition function. 
\end{titlepage}
\newpage

\renewcommand{\thefootnote}{\arabic{footnote}}
\setcounter{footnote}{0}

\section{Introduction}

Supersymmetric gauge theory is one of the most exciting 
topics in high energy physics from various points of view. 
An important property of supersymmetric theory is that 
we can exactly deal with the theory 
using the algebra of supersymmetry. 
Among such exact results, 
one of the most significant steps toward understanding 
the non-perturbative property of (supersymmetric) gauge theory 
is that Seiberg and Witten exactly determined 
the low energy prepotential of four-dimensional $\cN=2$ supersymmetric 
gauge theory \cite{Seiberg:1994rs,Seiberg:1994aj}. 
Furthermore, 
the instanton contribution to the prepotential of the $\cN=2$ theory 
has been exactly calculated through a partition function 
of Young diagrams (the Nekrasov partition function)
\cite{Nekrasov:2002qd,Losev:2003py}. 
This partition function 
is obtained by explicitly carrying out the path-integral 
of four-dimensional $\cN=2$ supersymmetric Yang-Mills theory 
deformed by a constant graviphoton background ($\Omega$-background)
parametrized by $\e$.%
\footnote{
The $\Omega$-background is parametrized by two 
parameters $\e_1$ and $\e_2$ in general.
In this paper, however, we concentrate on the case of 
$\e_1=-\e_2=\e$. 
}
Thanks to this deformation, we can carry out the integration 
over the instanton moduli space using the so-called 
localization technique 
\cite{Witten:1992xu}--%
\nocite{Cordes:1995fc,Moore:1997dj}\cite{Moore:1998et}, 
which yields a partition function of 
Young diagrams 
(see \cite{Bruzzo:2002xf} for further development). 
In particular, the leading term of the free energy in 
the expansion in powers of $\e$ has been shown to coincide 
with the low energy prepotential of 
four-dimensional $\cN=2$ $SU(N)$ supersymmetric Yang-Mills theory 
\cite{Nekrasov:2003rj}--\nocite{Nakajima:2003pg}\cite{Nakajima:2003uh}
(see also \cite{Maeda:2004is}). 
Thus the $k$-instanton contribution to the 
prepotential can be explicitly read off by evaluating 
the partition function.

The Nekrasov partition function and the nature of the 
$\Omega$-background have been intensively studied 
in terms of string theory. 
In general, four-dimensional $\cN=2$ gauge theory 
can be realized in Type IIB superstring theory 
as an effective theory on fractional D3-branes,  
where instanton effects come from D(-1)-branes 
bound to the fractional D3-branes. 
In \cite{Billo:2006jm}, it was shown that the effective 
action of the fractional D3-D(-1) system in a RR-background 
coincides with the instanton effective action of 
four-dimensional $\cN=2$ theory in the $\Omega$-background. 
This means the RR-background is equivalent with 
the $\Omega$-background in this brane configuration. 
This brane configuration in the RR-background and 
the deformed four-dimensional gauge theory have been 
further studied in 
\cite{Blumenhagen:2007bn}--%
\nocite{Ito:2007hy,Billo:2007sw,Sasaki:2007iv,Billo':2008sp}%
\cite{Billo':2008pg}. 
For a relation between the Nekrasov formula and 
topological string, see 
\cite{Iqbal:2003ix}--\nocite{Iqbal:2003zz,Eguchi:2003sj,%
Eguchi:2003it,Hollowood:2003cv}\cite{Zhou:2003zp}. 
In this connection, relations with 
topological vertex \cite{Aganagic:2002qg}--%
\nocite{Iqbal:2002we}\cite{Aganagic:2003db}, 
melting crystal \cite{Okounkov:2003sp,Iqbal:2003ds}, 
and integrable systems have been studied in detail 
\cite{Tachikawa:2004ur}--%
\nocite{Awata:2005fa,%
Matsuura:2005sg,%
Maeda:2005qg,%
Marshakov:2006ii,%
Maeda:2006we,%
Iqbal:2007ii,%
Marshakov:2007rh,%
Taki:2007dh,%
Nakatsu:2007dk,%
Marshakov:2007ra,%
Awata:2008ed,%
Nakatsu:2008mr}%
\cite{Nakatsu:2008px}%
\footnote{
For relations between the Nekrasov partition function 
and simple physical systems like 2D Yang-Mills theory, 
3D Chern-Simons theory and matrix models, 
see \cite{deHaro:2004id}--\nocite{Matsuo:2004cq}%
\cite{Tai:2007vc}.}.

The purpose of this paper is to reproduce the 
Nekrasov partition function in terms of the Hanany-Witten
type brane configuration in Type IIA superstring theory 
\cite{Hanany:1997ie,Witten:1997sc}, that is, 
a system of $N$ D4-branes stretched between two parallel NS5-branes. 
This configuration is a T-dual of the system of 
fractional D3-branes mentioned above  
and the instanton effects of four-dimensional Yang-Mills theory 
come from (Euclidean) D0-branes ``propagating'' 
between the two NS5-branes. 
We explicitly show that we can identify the $\Omega$-background as 
a background of RR 4-form field strength in this brane configuration 
as expected. 
In general, D0-branes behave as a dielectric D2-brane 
in a background of constant RR 4-form field strength
\cite{Myers:1999ps}. 
In addition, a D2-brane can have the end on a Type IIA NS5-brane  
whose boundary is coupled with the self-dual 2-form potential 
in the world-volume of the NS5-brane \cite{Strominger:1995ac}. 
We evaluate the potential energy coming from the interaction between 
the boundaries of the dielectric D2-branes 
in the NS5-branes and the ``kinetic energy'' of the D0-branes 
propagating between the two NS5-branes. 
We show that the partition function of these configurations  
coincides with the Nekrasov partition function of instantons. 
We also reproduce the Nekrasov partition function for 
$\cN=2$ theory with hypermultiplets in the fundamental 
representation using the same method.

The organization of this paper is as follows. 
In the next section, after briefly reviewing 
the Hanany-Witten type brane configuration, 
we show that the $\Omega$-background is 
identical with a background of RR 4-form field strength 
in this brane configuration. 
In Section 3, 
we analyze the behavior of D0-branes in detail 
and reproduce the Nekrasov partition function as a partition function 
of the D0-branes.
In Section 4, we reproduce the instanton partition function 
of $\cN=2$ theory with hypermultiplets in the fundamental representation 
using the same method developed in Section 3. 
Section 5 is devoted to conclusion and discussion. 
In Appendix A, we review the Frobenius representation of Young diagram. 
In Appendix B, we summarize the Nekrasov partition function 
and rewrite it in the Frobenius representation. 
In Appendix C, we summarize the Nekrasov formula for $\cN=2$ theory 
with hypermultiplets in the fundamental representation.

\section{Brane Configuration in a RR-background}

In this section, we briefly review the Hanany-Witten type brane 
configuration to realize four-dimensional $\cN=2$ supersymmetric 
gauge theory. 
We introduce NSNS B-field and RR 4-form field strength 
in the background and show that 
the Myers term introduced in the effective potential 
of D0-branes is identical with the deformation 
of instanton effective action 
by the $\Omega$-background.

\subsection{Brane configuration}

\begin{figure}[t]
\begin{center}
\includegraphics[scale=0.6]{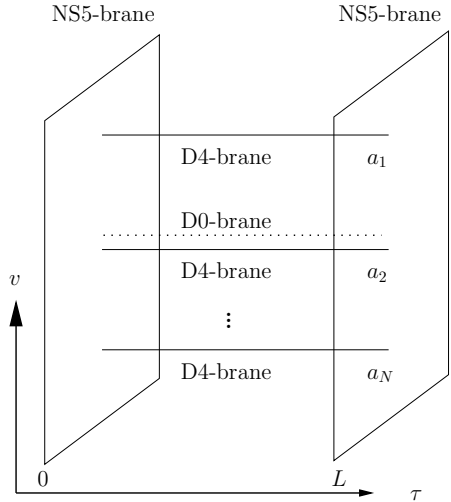}
\end{center}
\caption{
 The brane configuration in $(\tau, v, \bar{v})$ space. 
 The NS5-branes sit at $\tau=0$ and $\tau=L$, and the 
 D4-branes are expressed as solid lines 
 stretch between the NS5-branes at $v=a_1,\cdots,a_N$.
 We also consider Euclidean D0-branes propagating between 
 the NS5-branes, which is expressed as a dotted line in the figure. 
}
\label{branes fig}
\end{figure}

In order to realize four-dimensional $\cN=2$ supersymmetric 
$SU(N)$ Yang-Mills 
theory, we consider a system of two parallel NS5-branes 
and $N$ D4-branes stretching between the NS5-branes \cite{Witten:1997sc}
(see also the review \cite{Giveon:1998sr} and references therein). 
The world-volumes of the NS5-branes and the D4-branes 
are along the 012345 and 01236 directions, respectively, 
and the $\cN=2$ gauge theory arises as the low energy 
effective theory on the common directions 0123.
We introduce complex combinations of the coordinates as 
\begin{equation}
 z_1 \equiv x^0 + i x^1, \quad z_2 \equiv x^2 + i x^3, \quad 
  v \equiv x^4 + i x^5, 
\end{equation}
and rename $x^6$ as $\tau$ as well. 
We assume that the NS5-branes sit at $\tau=0$ and $L$, respectively, 
and the D4-branes stretch between NS5-branes at $v=a_l$ 
($l=1,\cdots,N$). 
Since we are interested in instanton contributions to the 
four-dimensional gauge theory, we also add (Euclidean) 
D0-branes bound to the D4-branes that propagate between 
the NS5-branes \cite{Losev:2003py} (see Fig.~\ref{branes fig}).
{}From the four-dimensional theory point of view, 
$a_l$ correspond to 
the classical vev of the adjoint scalar fields 
and the coupling constant of the gauge theory is related to $L$ as 
\begin{equation}
 \frac{1}{g^2_{\rm YM}} = \frac{L}{g_s l_s}, 
\label{gauge coupling}
\end{equation}
where $g_s$ and $l_s$ are the string coupling constant 
and the string length, respectively.

Precisely speaking, the definition of $L$ is ambiguous since 
the D4-branes deform the world-volume of the NS5-branes 
as \cite{Witten:1997sc} 
\begin{equation}
\tau = \pm g_s l_s \sum_{l=1}^N \Bigl(
  \log\frac{\left|v - a_l \right|}{\Lambda}
\Bigr),
\label{shape}
\end{equation}
where $\Lambda$ is a positive constant. 
From the relation (\ref{gauge coupling}), the gauge coupling constant
is a function of $v$. In the region $v \gg a_l$, it gives 
\begin{equation}
 \frac{1}{g_{\rm YM}^2(v)} \sim \log
  \frac{|v|^{2N}}{\Lambda^{2N}}. 
\end{equation}
Thus it is reasonable to interpret $\Lambda$ as the dynamical 
scale of the four-dimensional gauge theory 
and define $L$ as the distance between the NS5-branes 
at the cut-off scale.

Under this configuration, 
in addition to the coupling between light modes on D4-branes, 
there are couplings of these light modes 
to bulk gravity fields, 
to fields living on NS5-branes, 
and to massive modes on D4-branes in general. 
Since we are interested in gauge theory dynamics, 
these coupling must be small. 
To do so, we take the limit,  
\begin{equation}
g_s \to 0, \quad L/l_s \to 0.
\label{limit1}
\end{equation} 
{}Furthermore, in order for the system to be consistent with 
the interpretation of 
Coulomb branch of the four-dimensional gauge theory, the vev of 
the adjoint scalar must be sufficiently smaller than 
the Kaluza-Klein scale $1/L$. Therefore, $a_l$ must 
satisfy 
\begin{equation}
 a_{ln} \equiv a_l - a_n \ll l_s^2/L. 
 \label{limit2}
\end{equation}
We must also require 
\begin{equation}
 a_{ln} \ll l_s, \quad a_{ln} \ll L, 
 \label{limit3}
\end{equation}
in order to decouple massive modes of open strings on D4-branes.

\subsection{Effective action of D0-branes and its deformation by flux}

Suppose that $k$ D0-branes are bound to D4-branes and 
let us consider the low energy effective theory on the D0-branes. 
The degrees of freedom of the effective theory 
come from open strings between D0-branes 
and those between D0-branes and D4-branes. 
The former gives $k\times k$ bosonic complex matrices 
$B_1$, $B_2$ and $\phi$, 
which are collective coordinates corresponding to the directions 
$z_1$, $z_2$ and $v$, respectively,  
and fermionic complex matrices 
$\Psi_{B_1}$, $\Psi_{B_2}$ and $\eta$ with the same size. 
The latter gives bosonic complex matrices $I$ and $J$
with the size $k\times N$ and $N\times k$, respectively,  
and fermionic complex matrices $\Psi_I$ and $\Psi_J$
with the size $k\times N$ and $N\times k$, respectively.

The effective theory is a quantum mechanics of these 
matrix variables \cite{Berkooz:1996is,VanRaamsdonk:2001cg}. 
Since we are interested in BPS configuration of the system, 
however, it is sufficient to look at the potential terms, 
which is efficiently expressed as a BRST exact form as 
\begin{align}
 V=Q \Tr_k \Bigl(
  \chi_r \mu_r + \chi_c \mu_c +\chi_c^\dagger \mu_c^\dagger
  + \bphi \mu_{\eta}
 \Bigr), 
 \label{instanton potential}
\end{align}
with 
\begin{align}
 \mu_r &\equiv [B_1,B_1^\dagger] + [B_2,B_2^\dagger] 
  + II^\dagger - J^\dagger J, \nn \\
 \mu_c &\equiv [B_1,B_2] + IJ, \nn \\
 \mu_\eta &= [B_1,\Psi_{B_1}^\dagger] - [B_1^\dagger, \Psi_{B_1}]
  + [B_2,\Psi_{B_2}^\dagger] - [B_2^\dagger, \Psi_{B_2}]
  + \Psi_I I^\dagger + I\Psi_I^\dagger 
  - J^\dagger \Psi_J - \Psi_J^\dagger J,  
\label{moments}
\end{align}
where we have introduced auxiliary fermionic 
matrices $\chi_r$ and $\chi_c$ 
and their superpartners $H_r$ and $H_c$ with the size $k\times k$. 
The BRST transformation is given by 
\begin{alignat}{2}
 &Q B_i = \Psi_{B_i}, &\qquad &Q \Psi_{B_i} = [\phi, B_i], \quad (i=1,2)
 \nn \\
 &QI = \Psi_I, &\qquad &Q \Psi_I = \phi I - I a, \nn \\
 &QJ = \Psi_J, &\qquad &Q \Psi_J = -J\phi + a J, \nn \\
 &Q\chi_r = H_r, &\qquad &Q H_r = [\phi,\chi_r], \nn \\
 &Q\chi_c = H_c, &\qquad &Q H_c = [\phi,\chi_c], \nn \\
 &Q\bphi = \eta, &\qquad &Q\eta = [\phi,\bphi], \qquad Q\phi = 0, 
 \label{BRST trans}
\end{alignat}
where $a\equiv {\rm diag}(a_1,\cdots,a_N)$.
As a nature of a system of $k$ D0-branes, the effective theory 
possesses gauge symmetry $U(k)$. In fact, (\ref{instanton potential})
is invariant under the transformation, 
\begin{align}
 &B_i \to g B_i g^{-1}, \quad \phi \to g \phi g^{-1}, \quad
 I \to gI, \quad J \to Jg^{-1}, \nn \\
 &\Psi_{B_1} \to g \Psi_{B_i} g^{-1}, \quad \eta \to g \eta g^{-1}, 
 \quad
 \Psi_I \to g\Psi_I, \quad \Psi_J \to \Psi_J g^{-1},
 \label{gauge trans}
\end{align}
with $g\in U(k)$. 
Note that the potential (\ref{instanton potential}) is 
nothing but the instanton effective action of four-dimensional $\cN=2$ 
supersymmetric gauge theory.
Indeed the BPS configuration of D0-branes, that is, 
the instanton moduli space is determined by solving 
the ADHM equations, 
\begin{equation}
 \mu_r = 0, \qquad \mu_c = 0, 
  \label{ADHM eqs}
\end{equation}
as well as 
\begin{align}
 [\phi, B_i] = 0, \quad 
 \phi I - I a = 0, \quad 
 -J \phi + a J = 0, \quad 
[\phi,\bphi]=0. 
\end{align}
For more detail, see \cite{Dorey:2002ik} and references therein.

We deform the effective potential (\ref{instanton potential}) 
by introducing flux in the background of the brane configuration. 
We first introduce a constant NSNS B-field, 
\begin{equation}
 B^{(2)} = \frac{\zeta}{2} \left(
  dz_1\wedge d\bz_1 + dz_2\wedge d\bz_2\right). \qquad (\zeta > 0) 
 \label{B-field}
\end{equation}
It is familiar that NSNS B-field background introduces 
noncommutativity into the world-volume of D4-branes 
\cite{Seiberg:1999vs}; 
\begin{equation}
  [z_1,\bz_1] = \frac{\zeta}{2}, \quad 
  [z_2,\bz_2] = \frac{\zeta}{2}, 
\end{equation}
which modifies $\mu_r$ in the potential (\ref{instanton potential}) as 
\begin{equation}
 \mu_r \to \mu_r - \zeta. 
\label{deformation1}
\end{equation}
By this modification, it turns out that the size moduli of 
instantons cannot be zero, that is, the small instanton 
singularity of the moduli space is resolved 
\cite{Braden:1999zp,Nekrasov:1998ss}. 
{}From the D0-brane point of view, the presence of NSNS 
B-field prevents D0-branes to escape from D4-branes 
without breaking supersymmetry \cite{Seiberg:1999vs}.

In addition to the NSNS B-field, 
we further introduce a RR 3-form potential,  
\begin{equation}
 C^{(3)} = \e \left(v + \bar{v} \right)
  \left(-d\tau\wedge dz_1 \wedge d\bz_1 
  +d\tau\wedge dz_2 \wedge d\bz_2
  \right), 
\label{RR flux}
\end{equation}
or, equivalently, a RR 4-form field strength, 
\begin{equation}
 F^{(4)} = 2 \e \left(d\tau \wedge d x^4 \wedge dz_1 \wedge d\bz_1 
  -d\tau \wedge d x^4 \wedge dz_2 \wedge d\bz_2 
  \right). 
 \label{RR field strength}
\end{equation}
As shown in \cite{Myers:1999ps}, D0-branes have a coupling 
with RR 4-form field strength through the so-called Myers term,  
\begin{equation}
 F^{(4)}_{\tau ijk}\Tr\left(\Phi^i[\Phi^j,\Phi^k]\right),
  \label{Myers term}
\end{equation}
where $F^{(4)}$ is the 4-form field strength and 
$\Phi^i$ are the collective coordinates of D0-branes. 
From (\ref{RR field strength}) and (\ref{Myers term}), we see that 
the Myers term, 
\begin{equation}
 \e \Tr_k \Bigl\{
  (\phi + \bphi)\left([B_1,B_1^\dagger] - [B_2,B_2^\dagger]\right)
\Bigr\}, 
\end{equation}
is added to the effective potential (\ref{instanton potential}) 
due to the RR 4-form field strength (\ref{RR field strength}). 
It is easy to show that this modification is achieved by 
modifying the BRST charge $Q$ in 
(\ref{instanton potential}) to $Q_\e$ defined by 
\begin{align}
 Q_\e \Psi_{B_1} &= [\phi,B_1] + \e B_1,  \nn \\
 Q_\e \Psi_{B_2} &= [\phi,B_2] - \e B_2, 
 \label{modified Q}
\end{align}
and the others are the same with (\ref{BRST trans}). 
This is exactly the same deformation of the instanton 
effective action in the $\Omega$-background 
\cite{Nekrasov:2002qd}. 
Thus, as expected,  
we can conclude that the $\Omega$-background is equivalent 
with the background of RR 4-form field strength 
(\ref{RR field strength})
in the brane configuration given above.

By combining both the effects of NSNS B-field (\ref{B-field}) 
and RR 3-form (\ref{RR flux}), we obtain the deformed 
D0-brane effective potential;
\begin{equation}
 V_{\rm mod} = Q_\e \Tr \Bigl(
\chi_r (\mu_r - \zeta) + \chi_c \mu_c +\chi_c^\dagger \mu_c^\dagger
  + \bphi \mu_{\eta}
\Bigr), 
\label{modified potential}
\end{equation}
with (\ref{moments}) and (\ref{modified Q}). 


\section{Nekrasov partition function from D0-branes}

The instanton part of the Nekrasov partition function 
(\ref{inst partition}) is originally obtained by 
explicitly evaluating the integral, 
\begin{equation}
 Z_k(\e,\Lambda) = \int [dB_1 dB_2 \cdots] e^{-V_{\rm mod}}, 
\end{equation}
by using a property that the integral is localized at 
$Q_\e$-invariant points in the moduli space of instantons 
\cite{Moore:1997dj,Moore:1998et,Bruzzo:2002xf}. 
In the language of brane configuration, it corresponds 
to counting the BPS configurations of D0-D4 bound state 
in the background of the NSNS B-field (\ref{B-field}) 
and the RR 4-form field strength (\ref{RR field strength}). 
Such a configuration is obtained by solving the equations, 
\begin{align}
 &[B_1,B_1^\dagger] + [B_2,B_2^\dagger] 
 + II^\dagger - J^\dagger J = \zeta,  \nn \\
 &[B_1,B_2] + IJ=0, \nn \\
 &[B_1, \phi] = \e B_1, \qquad [B_2, \phi] = -\e B_2, \qquad 
 [\phi, \bphi] = 0, \nn \\ 
 &\phi I - I a = 0, \qquad J\phi - a J = 0. 
 \label{D0 equation}
\end{align}
In this section, we reproduce the instanton part of the Nekrasov 
partition function as a partition function of the D0-branes 
in the brane configuration introduced in the previous section. 
For simplicity, we start with the case of $N=1$
and the case of $N>1$ follows that. 

\subsection{$N=1$}

Although we can set $a=0$ in this case, 
we keep it for a later discussion. 
It is easy to show that the matrices, 
\begin{align}
 B_1^{(\tn,n;a)} &\equiv \left(
\begin{matrix}
 0 & \sqrt{\zeta} &        & & & & \\
   & \ddots   & \ddots & & & & \\
   &          & 0      & \sqrt{\tn\zeta} & & & \\
   &          &        & 0 & 0 & & \\
   &          &        &   & 0 & \ddots & \\
   &          &        &   &   & \ddots & 0 \\
   &          &        &   &   &        & 0 
\end{matrix}
\right), \quad 
 B_2^{(\tn,n;a)} = \left(
\begin{matrix}
 0 &        & & & & & \\
 0 & \ddots & & & & & \\
   & \ddots & 0 & & & & \\
   &        & 0 & 0 & & & \\
   &        &   & \sqrt{n\zeta} & 0 & & \\
   &        &   &          & \ddots & \ddots &   \\
   &        &   &          &        & \sqrt{\zeta} & 0 \\
\end{matrix}
\right), \nn \\
\phi^{(\tn,n;a)} &\equiv
\left(
\begin{matrix}
 a-\tn\e &       & & & & & \\
        &\ddots &    & & & & \\
        &       &a-\e& & & & \\
        &       &  &a&  & & \\
        &       &  & &a+\e& & \\
        &       &  & &  &\ddots& \\
        &       &  & &  & & a+n\e  
\end{matrix}
\right), \quad 
I^{(\tn,n;a)} \equiv 
\left(
\begin{matrix}
 0 \\ \vdots \\ 0 \\ \sqrt{(n+\tn+1)\zeta} \\ 0 \\ \vdots \\ 0 
\end{matrix}
\right), \nn \\ 
J^{(\tn,n;a)}&\equiv0, 
\label{irr sln}
\end{align}
solve the equations (\ref{D0 equation}) for any $\tn,n\in \Z_{\ge 0}$ 
with $\tn+n+1=k$. We call (\ref{irr sln}) as an irreducible solution.   
The equations (\ref{D0 equation}) can be generally solved 
as block diagonal matrices that consist of irreducible solutions;
\begin{align}
 B_i^{(\tilde\bfn,\bfn;a)} &\equiv 
\left(
\begin{matrix}
 B_i^{(\tn_1,n_1;a)} & & \\
 & \ddots & & \\
 & & B_i^{(\tn_L,n_L;a)}
\end{matrix}
\right), \quad 
 \phi^{(\tilde\bfn,\bfn;a)} \equiv 
\left(
\begin{matrix}
 \phi_i^{(\tn_1,n_1;a)} & & \\
 & \ddots & & \\
 & & \phi_i^{(\tn_L,n_L;a)}
\end{matrix}
\right), \nn \\
I^{(\tilde\bfn,\bfn;a)} &\equiv 
\left(
\begin{matrix}
 I^{(\tn_1,n_1;a)} \\ \vdots \\ I^{(\tn_L,n_L;a)}
\end{matrix}
\right), \quad J^{(\tilde\bfn,\bfn;a)} \equiv 0, 
\label{N=1 general sln}
\end{align}
up to the gauge transformation (\ref{gauge trans}) and/or 
a discrete transformation that keeps the potential 
(\ref{modified potential}) invariant. 
Here we have defined $\tilde\bfn$ and $\bfn$ as 
$\tilde\bfn=(\tn_1,\cdots,\tn_L)$ and $\bfn=(n_1,\cdots,n_L)$
with satisfying 
\begin{equation}
\sum_{i=1}^L(\tn_i+n_i+1)\equiv \sum_{i=1}^L k_i = k,
\end{equation} 
and we call this solution as a reducible solution in the following. 
Note that, by using the center $Z_k$ of the gauge group $U(k)$,
we can always rearrange $\tilde\bfn$ and $\bfn$ so that 
they satisfy 
\begin{equation}
 \tn_1 \ge \cdots \ge \tn_L \ge 0, \qquad 
  n_1 \ge \cdots \ge n_L \ge 0. 
\label{temporary condition}
\end{equation}

\begin{figure}[t]
\begin{center}
\includegraphics[scale=0.5]{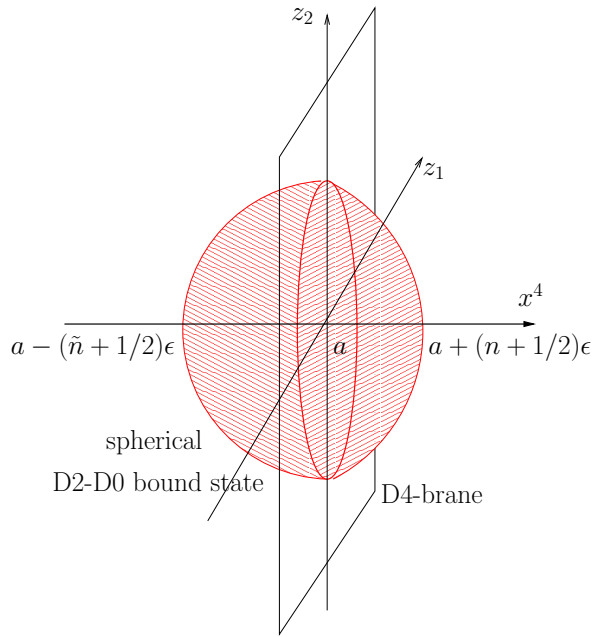}
\end{center}
\caption{
The spherical brane configuration corresponding to 
the irreducible solution (\ref{irr sln}). 
We can interpret this configuration as 
a fuzzy distribution of D0-branes or 
a spherical D2-brane to which D0-branes are bound. 
In the latter interpretation, the D0-branes locate at 
$x^4=a-\tn \e, a-(\tn-1)\e, \cdots, a+n\e$ in 
the spherical D2-brane, which cuts 
the $x^4$-axis at $x^4=a-(\tn+1/2)\e$ and 
$x^4=a+(n+1/2)\e$.  
}
\label{spherical D2 fig}
\end{figure}

What kind of configuration does the solution 
(\ref{N=1 general sln}) express? 
In order to answer this question, let us first recall 
the situation of D0-D4 bound state in a constant self-dual 
NSNS B-field background. 
From the view point of D4-brane, D0-branes behave as 
instantons of four-dimensional gauge theory whose size 
is bounded from below as a result of non-commutativity 
introduced by the B-field. 
As a result, the D0-branes are smeared into the four-dimensional 
(Euclidean) world-volume of the D4-branes with the size 
$\sqrt{|B|}$ \cite{Braden:1999zp}. 
On the other hand, consider D0-branes in a 
constant RR 4-form field strength background. 
In this case, the D0-branes expand 
into a fuzzy two-sphere, which can also be regarded 
as a spherical D2-brane to which D0-branes are bound 
\cite{Myers:1999ps}. 
Combining them, we can think of 
the irreducible solution (\ref{irr sln}) as a spherical 
configuration of D0-branes that is smeared in the four-dimensional 
space along $(z_1,z_2)$ and is bulging from 
the world-volume of D4-brane in the $x^4$-direction. 
Looking at the solution (\ref{irr sln}), 
the typical size of this configuration along 
$(z_1,z_2)$ is $\cO(\sqrt{\zeta})$ 
and that along the $x^4$-direction is $\cO(\e)$ as expected. 
This configuration can be also interpreted as a spherical 
D2-branes to which D0-branes are bound as mentioned above. 
From the irreducible solution $\phi^{(\tn,n;a)}$, we see that 
the D0-branes locate at 
$x^4=a-\tn \e, a-(\tn-1)\e, \cdots, a+n \e$. 
Since the D0-branes are bound to a spherical D2-brane, 
it is reasonable to think that the D2-brane 
cuts the $x^4$-axis at $x^4=-(\tn+1/2)\e$ and $x^4=(n+1/2)\e$ 
(Fig.~\ref{spherical D2 fig}). 
We can give the same interpretation for the reducible solution
(\ref{N=1 general sln}); it would express a set of 
$L$ spherical D2-branes that cut the $x^4$-axis at 
$x^4=a-(\tn_i+1/2)\e$ and $x^4=a+(n_i+1/2)\e$ 
to which D0-branes are bound at $x^4=a-\tn_i \e,\cdots,a+n_i \e$ 
$(i=1,\cdots,L)$.%
\footnote{
This is essentially the same configuration obtained in 
 \cite{Ito:2007hy}.  
} 
Note that the net D2-brane charge of this system is zero.  
This corresponds to the observation that we need two patches 
to make a 2-sphere, which are regarded as a D2-brane 
and an anti-D2-brane, respectively.

\begin{figure}[t]
\begin{center}
\includegraphics[scale=.7]{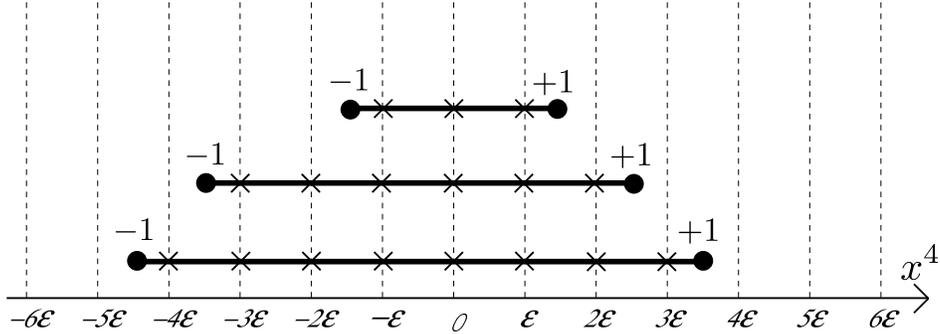}
\end{center}
\caption{
The boundary of the spherical D2-branes in the $v$-plane 
of the NS5-brane. We have depicted the case of 
$(\tilde\bfn|\bfn)=(4,3,1|3,2,1)$ and $a=0$ as an example. 
In the $v$-plane, the boundary of each spherical brane 
can be seen as a 1-dimensional object. 
The edges of the object have opposite charges that 
make Coulomb potential in the $v$-plane. 
The crosses on the objects express the positions of D0-branes. 
Although we have drawn each objects as if they were shifted 
in some direction, 
they are overlapping on the $x^4$-axis in reality.  
}
\label{bit strings fig}
\end{figure}

Since we are interested in dynamics of 
four-dimensional gauge theory and $\e$ must be much smaller than 
$a_{ln}$ (in the case of $N>1$), $\e$ and $\zeta$ should 
satisfy, 
\begin{equation}
\e \ll \sqrt{\zeta},  
\label{consistent limit}
\end{equation}
from the requirement of (\ref{limit2}) and (\ref{limit3}). 
In this limit, each spherical D2-brane 
behaves as a pair of D2-brane and anti-D2-brane at 
$x^4=-(\tn_i+1/2)$ and $x^4=n_i+1/2$, respectively, 
which are smeared in the four-dimensional space $(z_1,z_2)$.
In general, 
a D2-brane can end on a Type IIA NS5-brane and the boundary 
behaves as a ``string'' coupled with self-dual 2-form 
potential in the world-volume of NS5-brane \cite{Strominger:1995ac}. 
Recalling that the D0-branes propagate between the NS5-branes 
and the spherical D2-branes are smeared in a four-dimensional 
space $(z_1,z_2)$, 
the boundaries of the D2-branes and anti-D2-branes are also 
smeared in $(z_1,z_2)$, 
that would create a two-dimensional Coulomb potential 
in the $v$-plane that is transverse to $(z_1,z_2)$ 
in each of the two NS5-branes. 
Therefore, looking at the $v$-plane in one of the NS5-branes,  
the boundary of each spherical brane can be seen as a 1-dimensional object 
spanned from $x^4=a-(\tn_i+1/2)\e$ to $x^4=a+(n_i+1/2)\e$ whose edges 
have opposite charges $\pm 1$ for the two-dimensional Coulomb force
as long as (\ref{consistent limit}) is satisfied. 
We depict an example of the configuration in the $v$-plane 
in Fig.~\ref{bit strings fig}.

Now let us estimate the energy of the system corresponding 
to the solution (\ref{N=1 general sln}) coming from the boundaries. 
We must take into account 
(1) the Coulomb force between charges in the NS5-branes 
and 
(2) the ``kinetic energy'' of D0-branes 
propagating between the NS5-branes:

\vspace{.5cm}
\noindent
{\bf (1) Coulomb force between charges in the NS5-branes }

In each NS5-brane, we assign charges $-1$ and $+1$ to the edges at 
$v=a - \e(\tn_i+1/2)$ and 
$v=a + \e(n_i+1/2)$ $(i=1,\cdots,L)$, respectively.%
\footnote{
Precisely speaking, the charges of the edges are opposite 
in each of the NS5-branes. But we do not need to distinguish
them since the NS5-branes are separated with each other. 
} 
The potential energy created by the two-dimensional Coulomb force 
between a charge $q=\pm1$ at $v=x$ and a charge $q'=\pm1$ at $v=y$ 
can be written as%
\footnote{
Considering a M5-brane on which a D2-brane ends in M-theory, 
we can estimate the coefficient of (\ref{Coulomb potential}) 
using the M5 and M2 charges $\mu_{\rm M5}$ and $\mu_{\rm M2}$ as 
$\frac{\mu_{\rm M2}^2}{2\pi\mu_{\rm M5}}=1$. 
} 
\begin{equation}
 V(x,q;y,q') = - q q' \log \left|x-y\right|. 
\label{Coulomb potential}
\end{equation}
Therefore, the potential energies between the edges 
of the 1-dimensional objects can be estimated as  
\begin{equation}
 V_{1} = 2 \log \Biggl(
\e^{L}\frac{\prod_{i,j=1}^L \left|n_i+\tn_j+1\right|}
{\prod_{i<j}^L\left|n_i-n_j\right|\left|\tn_i-\tn_j\right|}
\Biggr), 
\label{Coulomb contribution}
\end{equation}
where the factor $2$ comes from the same effect from the two 
NS5-branes. 

\vspace{.5cm}
\noindent
{\bf (2) D0-branes propagating between the NS5-branes}

We estimate the kinetic energy of each spherical brane 
by regarding that $k_i$ D0-branes at 
$x^4=a-\tn_i \e,\cdots,a+n_i \e$ are propagating from 
one NS5-brane to the other NS5-brane. 
From (\ref{shape}), we see that 
the distance between the NS5-branes at the position $v$ 
is given by 
\begin{equation}
 d(v) = 2 g_s l_s \Bigl(
\log \frac{|v-a|}{\Lambda} - \delta_{v,a}\log 0
\Bigr), 
\end{equation}
where the second term is necessary to regularize the distance 
at $v=a$ as $d(a)=-2 g_s l_s \log\Lambda$. 
By summing up all the contribution from the D0-branes, 
we can estimate the kinetic energy as 
\begin{equation}
 V_2 = \sum_{i=1}^L \sum_{m=-\tn_i}^{n_i} T_0'\, d(a+m\e) 
= 2 g_s l_s T_0' \log\Biggl(
\frac{\e^{k-L}}{\Lambda^k}
\prod_{i=1}^L n_i! \tn_i!
\Biggr), 
\label{D0 contribution}
\end{equation}
where $T_0'$ is the effective mass of the D0-brane. 

\vspace{.5cm}

Here we assume that we can use the mass of a single D0-brane 
as the effective mass $T_0'$: 
\begin{equation}
T_0'=\frac{1}{g_s l_s}. 
\label{coefficient assumption}
\end{equation}
Then we can write down the Boltzmann weight of this configuration 
as 
\begin{align}
 Z(\tilde{\bfn},\bfn,\e,\Lambda) 
&\equiv e^{-V_1-V_2}
 = \frac{\Lambda^{2k}}{\e^{2k}}
\frac
{\prod_{i<j}^L\left|n_i-n_j\right|^2\left|\tn_i-\tn_j\right|^2}
{\prod_{i,j=1}^L \left|n_i+\tn_j+1\right|^2}
\Biggl(
\frac{1}{\prod_{i=1}^L n_i! \tn_i! }
\Biggr)^2. 
\label{N=1 Boltzmann}
\end{align}
This is nothing but the Boltzmann weight of the Nekrasov partition 
function in the Frobenius representation 
(\ref{Nekrasov in Frobenius}) for $N=1$! 
Note that, if the configuration $(\tilde{\bfn},\bfn)$ satisfies 
$\tn_i=\tn_{i+1}$ or $n_i=n_{i+1}$ for some $i$, the corresponding 
Boltzmann weight becomes zero since the potential energy 
(\ref{Coulomb potential}) diverges. 
Therefore we can effectively require that $(\tilde{\bfn},\bfn)$ 
satisfy
\begin{equation}
 \tn_1 > \cdots > \tn_L \ge 0, \qquad 
  n_1 > \cdots > n_L \ge 0,  
\label{monotonic condition}
\end{equation}
instead of (\ref{temporary condition}). 

\subsection{$N>1$}

\begin{figure}[t]
\begin{center}
\includegraphics[scale=0.6]{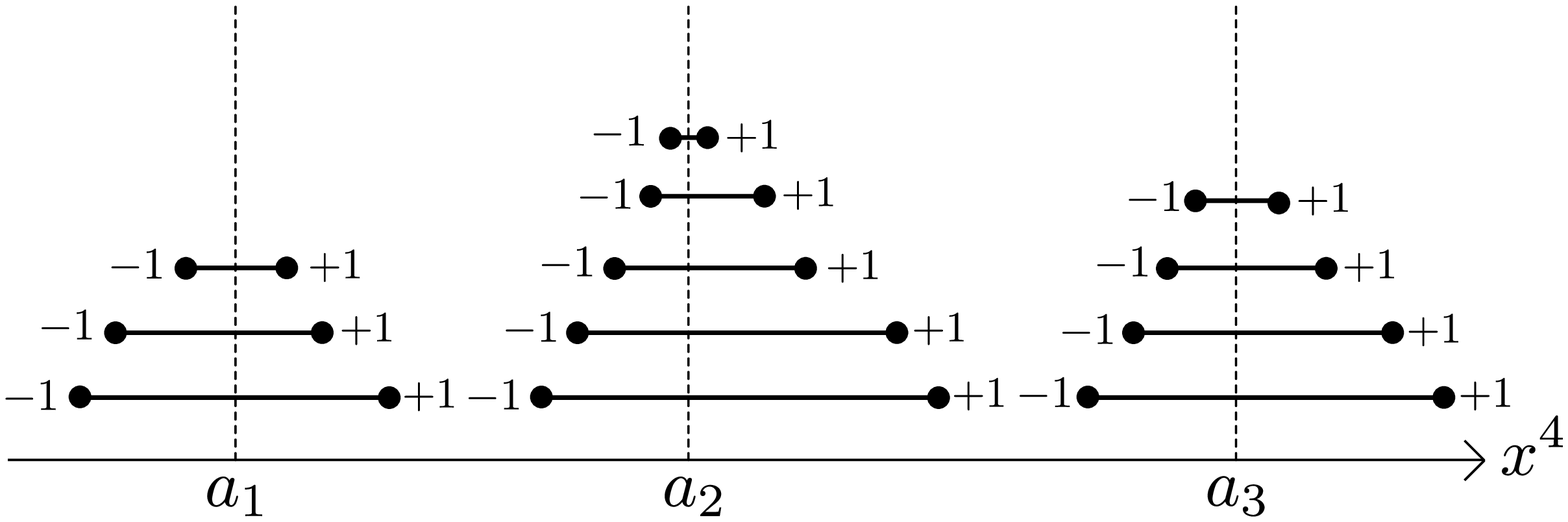}
\end{center}
\caption{
The configuration in the $v$-plane in 
the case of $N>1$. We have depicted the case of $N=3$  
and $(L_1,L_2,L_3)=(3,5,4)$ as an example. 
As same as the Fig.~\ref{bit strings fig}, 
at the edges of each 1-dimensional object have 
opposite charges that make a Coulomb potential 
in the $v$-plane. 
}
\label{general 2D config fig}
\end{figure}

It is straightforward to extend the above analysis to the 
case of $N>1$. 
We can write down a general solution of the equation 
(\ref{D0 equation})
using the reducible solution (\ref{N=1 general sln}); 
\begin{align}
 B_i^{(\tilde{\bfn}^l,\bfn^l,a_l)_{l=1}^N} 
&\equiv \bigoplus_{l=1}^N B_i^{(\tilde{\bfn}^l,\bfn^l,a_l)}, \quad 
\phi^{(\tilde{\bfn}^l,\bfn^l,a_l)_{l=1}^N} 
\equiv \bigoplus_{l=1}^N \phi^{(\tilde{\bfn}^l,\bfn^l,a_l)}, \nn \\  
I^{(\tilde{\bfn}^l,\bfn^l,a_l)_{l=1}^N}& \equiv 
\left(
\begin{matrix}
 I^{(\tilde{\bfn}_1,\bfn_1,a_1)} \\ \vdots \\ 
 I^{(\tilde{\bfn}_N,\bfn_N,a_N)}
\end{matrix}
\right), \quad 
J^{(\tilde{\bfn}^l,\bfn^l,a_l)_{l=1}^N} \equiv 0, 
\label{general sln}
\end{align}
where $\tilde{\bfn}^l=(\tn_1^l,\cdots,\tn_{L_l}^l)$ and 
$\bfn^l=(n_1^l,\cdots,n_{L_l}^l)$ again 
satisfy (\ref{temporary condition}) and 
\begin{equation}
 \sum_{l=1}^N \sum_{i=1}^{L_l} (\tn_i^l + n_i^l +1) \equiv 
 \sum_{l=1}^N k_l =k. 
\end{equation}
In terms of D-brane configuration, this solution 
corresponds to $L_l$ spherical D2-branes 
around $v=a_l$ $(l=1,\cdots,N)$; the positions of the $N$ D4-branes. 
Therefore, looking at this configuration in the $v$-plane, 
there seem to be $L_l$ 1-dimensional objects around $v=a_l$ 
$(l=1,\cdots,N)$, which have opposite charges at the edges 
$x^4=a_l- (\tn_i^l+1/2)\e$ and $x^4=a_l+ (n_i^l+1/2)\e$ $(i=1,\cdots,L_l)$ 
(see Fig.~\ref{general 2D config fig}). 

We again estimate the energy of this configuration 
by summing up the potential energy by the two-dimensional Coulomb 
force in the NS5-branes 
and the kinetic energy of the D0-branes propagating 
between the NS5-branes.  
It is easy to see that the potential energy for the Coulomb force 
is given by 
\begin{align}
 V_1 = 2 \log \Biggl\{
&\prod_{l=1}^{N}
\left(
\e^{L_l}
\frac
{\prod_{i,j}^{L_l}|n_i^{l}+\tilde{n}_j^{l}+1|}
{\prod_{i < j}^{L_l}|n_i^l-n_j^{l}|
|\tilde{n}_i^{l}-\tilde{n}_j^{l}|}
\right)
\nn \\
\times
&\prod_{l<n}^N \Biggl(
\prod_{i=1}^{L_l}
\prod_{j=1}^{L_n}
\frac
{|a_{ln}+\epsilon(n_i^{l}+\tilde{n}_j^{n}+1)|
|a_{ln}-\epsilon(\tilde{n}_i^{l}+n_j^{n}+1)|}
{|a_{ln}+\epsilon(n_i^{l}-n_j^{n})|
|a_{ln}-\epsilon(\tilde{n}_i^{l}-\tilde{n}_j^{n})|}
\Biggr)
\Biggr\}.
\label{general Coulomb}
\end{align}
On the other hand, since the distance between the NS5-branes 
is now given by 
\begin{equation}
 d(v) = 2 g_s l_s \sum_{l=1}^N \left(
\log\frac{|v-a_l|}{\Lambda} - \delta_{v,a_l} \log 0
\right), 
\end{equation}
the kinetic energy of the D0-branes becomes 
\begin{align}
 V_2 = 2g_s l_s T_0'\log\Biggl\{
&\prod_{l=1}^N\left(
\frac{\e^{k_l-L_l}}{\Lambda^{k_l}}
\prod_{i=1}^{L_l} \tn_i^l! n_i^l!
\right) \nn \\
&\times\prod_{l=1}^N \prod_{n\ne l}^N 
\prod_{i=1}^{L_l}
\left(
|a_{ln}-\e \tn_i^l|\cdot
|a_{ln}-\e (\tn_i^l-1)| \cdots |a_{ln}+\e n_i^l| 
\right)
\Biggr\}. 
\label{general kinetic}
\end{align}
From (\ref{general Coulomb}) and (\ref{general kinetic}), 
we obtain the Boltzmann weight of the D0-branes 
corresponding to the solution (\ref{general sln}):  
\begin{align}
Z(\tilde{\bfn}^l,\bfn^l,\bfa,\e,\Lambda)
&\equiv e^{-V_1-V_2} \nn \\
= \frac{\Lambda^{2kN}}{\e^{2kN}} 
&\prod_{l=1}^{N}
\left\{
\frac{\prod_{i < j}^{L_l}
|n_i^l-n_j^{l}|^2
|\tilde{n}_i^{l}-\tilde{n}_j^{l}|^2}{
\prod_{i,j}^{L_l}|n_i^{l}+\tilde{n}_j^{l}+1|^2
}
\left(
\prod_{i=1}^{L_l}
\frac{1}{
n_i^{l}!\tilde{n}_i^{l}!
}
\right)^2
\right\}
\nonumber\\
\times
&\prod_{l<n}^N\Biggl\{
\prod_{i=1}^{L_l}
\prod_{j=1}^{L_n}
\frac{
|a_{ln}+\epsilon(n_i^{l}-n_j^{n})|^2
|a_{ln}-\epsilon(\tilde{n}_i^{l}-\tilde{n}_j^{n})|^2
}{|a_{ln}+\epsilon(n_i^{l}+\tilde{n}_j^{n}+1)|^2
|a_{ln}-\epsilon(\tilde{n}_i^{l}+n_j^{n}+1)|^2
}
\nonumber\\
&\hspace{1cm}\times \prod_{i=1}^{L_l}
\frac{1}{
|a_{ln}-\epsilon \tilde{n}_i^{l} |^2\cdot
|a_{ln}-\epsilon (\tilde{n}_i^{l}-1)|^2
\cdots 
|a_{ln}+\epsilon n_i^{l} |^2
}\nn \\
&\hspace{1cm}\times
\prod_{j=1}^{L_n}
\frac{1}{
{|a_{ln}+\epsilon \tilde{n}_j^{n}|^2\cdot
|a_{ln}+\epsilon (\tilde{n}_j^{n}-1)|^2
\cdots
|a_{ln}-\epsilon n_j^{n}|^2}
}
\Biggr\}, 
\label{D0 partition N>1}
\end{align}
under the assumption (\ref{coefficient assumption}). 
This expression again 
coincides with the instanton part of the Nekrasov partition 
function in the Frobenius representation (\ref{Nekrasov in Frobenius}). 
From this result, 
we conclude that the Nekrasov partition function is 
that of D0-branes bound to D4-branes in the presence of 
NSNS B-field (\ref{B-field}) and RR 3-form (\ref{RR flux})
in the background of the Hanany-Witten type brane configuration.

\section{Nekrasov partition function for ${\cal N}=2$ QCD from D0-branes}

In this section, we reproduce the Nekrasov partition function 
for four-dimensional $\cN=2$ theory with hypermultiplets in the fundamental 
representations \cite{Nekrasov:2002qd} 
as a non-trivial check of 
the method developed in the previous section.

\begin{figure}[t]
\begin{center}
\includegraphics[scale=0.6]{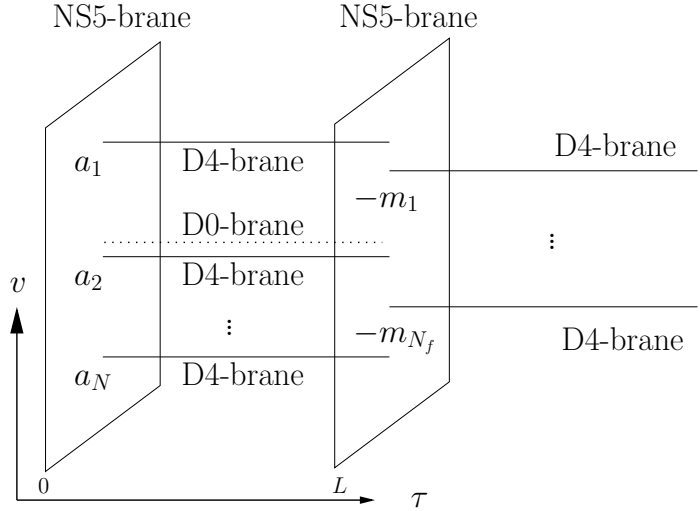}
\end{center}
\caption{
 The brane configuration to realize four-dimensional $\cN=2$ supersymmetric 
 gauge theory with hypermultiplets in the fundamental representation. 
 In addition to Fig.~\ref{branes fig}, we have added
 $N_f$ D4-branes that attach to one of the NS5-branes 
 at $v=-m_1,\cdots,-m_{N_f}$. They have semi-infinite world-volumes 
 in the direction $\tau$. 
}
\label{fundamental fig}
\end{figure}

In order to introduce matter fields in the fundamental representation, 
we add $N_f$ semi-infinite D4-branes attached to one of 
the NS5-branes at $v=-m_1, \cdots, -m_{N_f}$ 
in addition to the $N$ D4-branes stretched between the 
NS5-branes (Fig.~\ref{fundamental fig}). 
The positions of these D4-branes correspond 
to the bare masses of the hypermultiplets. 

The strategy to construct the partition function of D0-branes 
is the same with the previous section, namely 
we use as the same configuration of D0-branes 
$\{(\tilde{\bfn}_l,\bfn_l,a_l)\,|\, l=1,\cdots,N\}$
as is used in the previous section. 
Then the potential energy for the Coulomb force in the $v$-plane 
is the same with (\ref{general Coulomb}). 
However, the kinetic energy is different from (\ref{general kinetic}) 
since the NS5-branes are further deformed 
from (\ref{shape}) by the presence 
of additional D4-branes:
\begin{align}
 \tau_{-} &= -g_s l_s \sum_{l=1}^N 
\log\Bigl(\frac{|v-a_l|}{\Lambda}\Bigr),  \nn \\
 \tau_+ &= g_s l_s\Biggl[
  \sum_{l=1}^N \log\Bigl(\frac{|v-a_l|}{\Lambda}\Bigr)
 -\sum_{f=1}^{N_f} \log\Bigl(\frac{|v+m_f|}{\Lambda}\Bigr)
\Biggr], 
\label{shapes2}
\end{align}
where $\tau_{\pm}$ are the positions of the NS5-branes 
in the direction $\tau$. 
Then, we see that the distance between 
the the NS5-branes at $v$ is given by 
\begin{equation}
 d(v)=g_s l_s\Biggl[
  2\sum_{l=1}^N \log\Bigl(\frac{|v-a_l|}{\Lambda}\Bigr)
 -\sum_{f=1}^{N_f} \log\Bigl(\frac{|v+m_f|}{\Lambda}\Bigr)
\Biggr], 
\end{equation}
and the kinetic energy of the D0-branes can be estimated as 
\begin{align}
 V_2 = g_s l_s T_0'\log\Biggl\{
&\prod_{l=1}^N\left(
\frac{\e^{k_l-L_l}}{\Lambda^{k_l}}
\prod_{i=1}^{L_l} \tn_i^l! n_i^l!
\right)^2
\times 
\prod_{l=1}^N \prod_{n\ne l}^N 
\prod_{i=1}^{L_l}
\Bigl(
|a_{ln}-\e \tn_i^l| \cdots |a_{ln}+\e n_i^l| 
\Bigr)^2 \nn \\
&\times
\Lambda^{k N_f}\prod_{f=1}^{N_f}\prod_{l=1}^N\prod_{i=1}^{L_l}
\Bigl(
\frac{1}{|m_f+a_l-\e\tn_i^l|\cdots|m_f+a_l+\e n_i^l|}
\Bigr)
\Biggr\}. 
\label{fundamental kinetic}
\end{align}
Combining (\ref{general Coulomb}) and (\ref{fundamental kinetic}) 
and assuming (\ref{coefficient assumption}), 
we obtain the Boltzmann weight corresponding to this configuration 
of D0-branes; 
\begin{align}
 Z(\tilde{\bfn}^l,\bfn^l,\bfa,\bfm,\e,\Lambda)= 
 &Z(\tilde{\bfn}^l,\bfn^l,\bfa,\e,\Lambda) \nn \\
 &\times 
 \frac{1}{\Lambda^{kN_f}}
 \prod_{f=1}^{N_f}\prod_{l=1}^N\prod_{i=1}^{L_l}
\Bigl(
{|m_f+a_l-\e\tn_i^l|\cdots|m_f+a_l+\e n_i^l|}
\Bigr), 
\label{Boltzman fundamental}
\end{align}
where $Z(\tilde{\bfn}^l,\bfn^l,\bfa,\e,\Lambda)$ is given by 
(\ref{D0 partition N>1}).
Looking at (\ref{Nek for fund in Frob}), 
we see that (\ref{Boltzman fundamental}) 
coincides with the instanton part of the Nekrasov partition function 
for four-dimensional $\cN=2$ theory with $N_f$ hypermultiplets 
in the fundamental representation.

\section{Conclusion and Discussion}

In this paper, we analyzed the behavior of D0-branes 
in the Hanany-Witten type brane configuration 
in a background of RR 4-form field strength 
and NSNS B-field. 
We showed that the partition function of Euclidean D0-branes 
propagating between the NS5-branes coincides with 
the Nekrasov partition function of instantons 
in four-dimensional $\cN=2$ supersymmetric Yang-Mills theory. 
In this analysis, the Myers effect played an important role. 
We applied the same method to the brane configuration 
realizing four-dimensional $\cN=2$ theory QCD 
and the partition function of the D0-branes again 
coincides with the Nekrasov partition 
function of the theory.

There would be many applications in the method 
developed in this paper. 
As a straightforward application, we can apply it 
to $\cN=2$ quiver gauge theories and/or $\cN=2$ theories 
with other gauge groups than $SU(N)$ \cite{LM}. 
It would also be interesting to reduce supersymmetry 
from $\cN=2$ to $\cN=1$ by deforming the NS5-branes 
holomorphically, which might give a connection to 
Dijkgraaf-Vafa theory \cite{Dijkgraaf:2002fc}%
--\nocite{Dijkgraaf:2002vw}\cite{Dijkgraaf:2002dh}. 
Although we have concentrated on four-dimensional theories 
in this paper, a five-dimensional version of the 
instanton partition function is also proposed \cite{Nekrasov:2002qd}. 
In terms of the brane configuration we have used in this paper, 
this would be achieved by lifting it up to a configuration of a M5-brane 
in the background of a 4-form field strength in the M-theory. 
It is interesting to extend our analysis to the M-theory 
and see how the five-dimensional version of the instanton 
partition function appears.

Lastly, as mentioned in Introduction,
the Nekrasov partition function 
is known to be equivalent 
to amplitudes of topological string 
theory of local toric Calabi-Yau manifolds 
\cite{Iqbal:2003ix}--\nocite{Iqbal:2003zz,Eguchi:2003sj,%
Eguchi:2003it,Hollowood:2003cv}\cite{Zhou:2003zp}. 
In the heart of this relation, there is an idea of 
geometrical engineering \cite{Katz:1996fh,Katz:1997eq}; 
by realizing 4D $\cN=2$ gauge theory by compactifying Type II 
superstring theory by a Calabi-Yau three-fold, 
some nature of the gauge theory is explained as a geometrical 
property of the Calabi-Yau manifold. 
It is interesting that the same partition function is 
obtained from rather simple set-up of branes in 
a RR-background. 
From this result, it would be quite natural 
to expect that the brane system 
with RR flux would be connected to 
a Calabi-Yau set-up by a sequence of string duality. 
It would be an important and interesting future work 
to reveal this connection.

\section*{Acknowledgments}
The author would thank 
T.~Asakawa,
R.~Boels,
M.~Hanada,
S.~Hirano,
K.~J.~Larsen,
and
K.~Zoubos
for useful discussion.
He would also thank P.~H.~Damgaard and N.~Obers 
for valuable comments and careful reading of this manuscript. 
This work is supported 
by JSPS Postdoctoral Fellowship for Research Abroad.

\appendix

\section{Frobenius Representation of Young diagram}

In this appendix, we introduce the Frobenius representation 
of Young diagram. 

We start with a Young diagram parametrized by $\{k_i\}$, 
the number of boxes in the $i$'s row satisfying 
\begin{gather}
 k_1 + \cdots + k_r = k, \qquad 
 k_1 \ge k_2 \ge \cdots \ge k_r > 0, 
\label{partition}
\end{gather}
which is a partition of the integer $k$. 
So we can identify the Young diagram with the partition 
$\bfk=\{k_1,\cdots,k_r\}$ itself.

\begin{figure}[t]
\begin{center}
\scalebox{.8}{\includegraphics{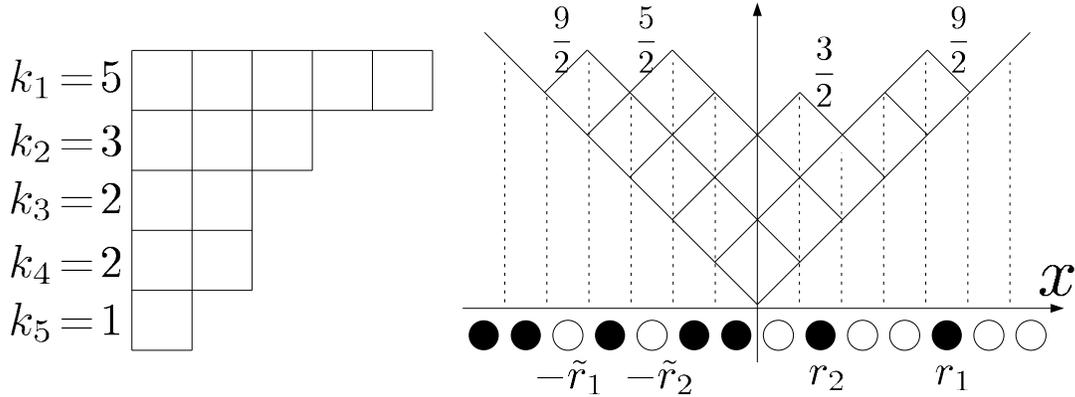}}
\end{center}
\caption{
An example of Young diagram in the Frobenius 
representation. 
The Young diagram in the left figure is expressed 
by the partition $\bfk=\{5,3,2,2,1\}$. 
The right figure is the corresponding profile function. 
We see that the Frobenius representation of this Young diagram is 
given by $(9/2,5/2\,|\,9/2,3/2)$. 
We also see that there are white circles at $x=-9/2$ and $-5/2$ 
and there are black circles at $x=9/2$ and $3/2$, which is 
the corresponding Maya diagram. 
}
\label{Young diagram fig}
\end{figure} 

For our purpose to write down the Nekrasov partition function, 
it is useful to draw the diagram by inclining 45 degrees. 
Here we divide the (inclined) diagram into two parts 
by the center line (Fig.~\ref{Young diagram fig}). 
Suppose there are $L$ boxes on this line. 
Let $\tilde{r}_i$ and $r_i$ $(i=1,\cdots,L)$ 
denote the numbers of boxes 
in the left and right of the $i$'s box on the center, 
respectively. 
By counting the ``number'' of the center box 
in the left and right as $1/2$, respectively,  
we can express the Young diagram $\bfk$ by a set of 
half integers $\tilde{r}_i$ and $r_i$:
\begin{equation}
 (\tilde\bfr|\bfr)=(\tilde{r}_1,\cdots,\tilde{r}_L|r_1,\cdots,r_L), \qquad 
  \tilde{r}_i,\,r_i \in \N-1/2.
\label{Frobenius representation}
\end{equation}
This expression is called the Frobenius representation of 
the Young diagram (or the partition) $\bfk$. 

Incidentally, 
this representation is deeply related with the so-called Maya diagram, 
which is defined as a sequence of black and white circles on a line. 
We start the situation where the black circles and the white circles 
are at $x\in-\N + 1/2$ and $x \in \N-1/2$, respectively. 
Maya diagram is obtained by exchanging the positions of 
arbitrary pairs of black and white circles. 
By construction, the number of white circles in the region 
$x<0$ and that of black circles in the region $x>0$ 
is the same. 
In the Frobenius representation (\ref{Frobenius representation}), 
$-\tilde{r}_i$ and $r_i$ are identified with the positions 
of white circles in $x<0$ and black circles in $x>0$, respectively. 
We draw an example of these relations in Fig.~\ref{Young diagram fig}. 

\section{Nekrasov Partition Function in the Frobenius Representation}

In this appendix, we review the Nekrasov partition function 
(for $\e_1=-\e_2\equiv\e$) 
and rewrite it in the Frobenius representation. 

The Nekrasov partition function is given by 
\begin{align}
Z_{\rm Nek}\left(\av,\e,\Lambda \right) &= 
Z_{\rm pert}(\av,\e)\sum_{k=1}^\infty 
\sum_{\substack{
 k_1,\cdots,k_N\in \Z_{\ge 0} \\
 k_1+\cdots+k_N=k}}
\sum_{ \vk_1\in Y_{k_1}}
\cdots\sum_{\vk_N\in Y_{k_N}}
Z_{\rm inst}(\av,\bfk,\e,\Lambda), 
\label{Nekrasov partition}
\end{align}
with
\begin{align}
Z_{\rm pert}(\av,\e) &= \exp\left\{
\sum_{l \neq n}\gamma_\e(a_l-a_n;\Lambda)
\right\},\\
\label{inst partition}
Z_{\rm inst}(\av,\bfk,\e,\Lambda)
&=
\Lambda^{2 N k} 
\prod_{(l,i)\neq(n,j)}
\frac{a_{ln}+\e(k_{l,i}-k_{n,j}+j-i)}{a_{ln}+\e(j-i)}, 
\end{align}
where $i,j=1,\cdots,\infty$, 
$l,n=1,\cdots,N$, $a_{ln}\equiv a_l-a_n$, 
$\gamma_\e(x;\Lambda)$ is defined through
the deference equation, 
\begin{equation}
\gamma_\e(x+\e;\Lambda)+\gamma_\e(x-\e;\Lambda)
-2\gamma_\e(x;\Lambda)=\log(x/\Lambda),  
\label{gamma function}
\end{equation}
and $Y_{l}$ is a set of Young diagrams with $l$ boxes, 
whose element is given as a partition of $l$, namely 
$\vec{l}=(l_1,l_2,\cdots)$ satisfying $ l_1 + l_2 + \cdots = l$ 
with $l_1 \ge \cdots \ge l_r > l_{r+1} = l_{r+2} \cdots = 0$. 

As shown in \cite{Nekrasov:2003rj},
the Nekrasov partition function can be compactly expressed 
using a piecewise-linear function called 
the (colored) profile function, 
\begin{equation}
 f_{\bfa,\bfk}(x|\e) \equiv \sum_{l=1}^N f_{\bfk_l}(x-a_l|\e), 
\label{colored profile}
\end{equation}
with 
\begin{equation}
f_\bfk(x|\e) \equiv |x| + \sum_{i=1}^\infty \left[
|x-\e(k_i-i+1)| - |x - \e(k_i - i)) - |x-\e(-i+1)| + |x - \e(-i)|
\right], 
\label{extended profile}
\end{equation}
where $\bfk_l$ are Young diagrams sitting at $x=a_l$. 
Using (\ref{colored profile}), the instanton part of 
the Nekrasov partition function (\ref{inst partition})
can be written as 
\begin{equation}
 Z_{\rm inst}(\av,\bfk,\e,\Lambda) = \exp\Bigl(
-\frac{1}{4}\int_{x\ne y} dx dy 
f''_{\bfa,\bfk}(x|\e) f''_{\bfa,\bfk}(y|\e) \gamma_\e(x-y;\Lambda)
\Bigr). 
\label{Nekrasov by profile}
\end{equation}

We can rewrite the profile function 
(\ref{extended profile})
using the Frobenius representation 
(\ref{Frobenius representation}). 
To this end, we express the Young diagrams $\bfk_l$ by 
$(\tilde{\bfr}_l|\bfr_l)$ $(l=1,\cdots,N)$ and 
introduce the integers, 
\begin{align}
 \tn_i^l \equiv \tilde{r}_i^l-1/2, \quad 
 n_i^l \equiv r_i^l-1/2, \qquad (i=1,\cdots,L_l)
\end{align}
which satisfy 
\begin{equation}
 \tn_1^l > \cdots > \tn_{L_l}^l \ge 0, \qquad 
  n_1^l > \cdots > n_{L_l}^l \ge 0. 
\end{equation}
Using this, the second derivative of 
the profile function can be rewritten as 
\begin{equation}
 f_\bfk''(x|\e) = \frac{1}{2}\Bigl[
\delta(x) + \sum_{i=1}^L\Bigl(
-\delta(x+\e(\tn_i+1)) + \delta(x+\e\tn_i)
+\delta(x-\e n_i) - \delta(x-\e(n_i+1)) 
\Bigr)
\Bigr].  
\label{extended Frobenius profile}
\end{equation}
Substituting this expression into (\ref{Nekrasov by profile}), 
we can rewrite the instanton part 
of the partition function (\ref{inst partition}) as 
\begin{align}
Z_{\rm inst}&(\av,\bfk,\e,\Lambda)
= \frac{\Lambda^{2kN}}{\e^{2kN}} 
\prod_{l=1}^{N}
\left\{
\frac{\prod_{i < j}^{L_l}
|n_i^l-n_j^{l}|^2
|\tilde{n}_i^{l}-\tilde{n}_j^{l}|^2}{
\prod_{i,j}^{L_l}|n_i^{l}+\tilde{n}_j^{l}+1|^2
}
\left(
\prod_{i=1}^{L_l}
\frac{1}{
n_i^{l}!\tilde{n}_i^{l}!
}
\right)^2
\right\}
\nonumber\\
\times
&\prod_{l<n}^N \Biggl\{
\prod_{i=1}^{L_l}
\prod_{j=1}^{L_n}
\frac{
|a_{ln}+\epsilon(n_i^{l}-n_j^{n})|^2
|a_{ln}-\epsilon(\tilde{n}_i^{l}-\tilde{n}_j^{n})|^2
}{|a_{ln}+\epsilon(n_i^{l}+\tilde{n}_j^{n}+1)|^2
|a_{ln}-\epsilon(\tilde{n}_i^{l}+n_j^{n}+1)|^2
}
\nonumber\\
&\hspace{1cm}\times \prod_{i=1}^{L_l}
\frac{1}{
|a_{ln}-\epsilon \tilde{n}_i^{l} |^2
\cdots
|a_{ln}+\epsilon n_i^{l} |^2
}
\times
\prod_{j=1}^{L_n}
\frac{1}{
{|a_{ln}+\epsilon \tilde{n}_j^{n}|^2
\cdots
|a_{ln}-\epsilon n_j^{n}|^2}
}
\Biggr\}. 
\label{Nekrasov in Frobenius}
\end{align}

\section{Nekrasov Partition Function for Four-dimensional $\cN=2$ QCD}

In this appendix, we summarize the Nekrasov partition function 
for four-dimensional $\cN=2$ theory with hypermultiplets 
in the fundamental representation and rewrite it 
using the Frobenius representation. 

Let $\bfm$ denote the vector of bare masses of the hypermultiplets:
\begin{equation}
 \bfm = (m_1,\cdots,m_{N_f}). 
\end{equation}
Then the Nekrasov partition function is given by 
\cite{Nekrasov:2002qd}
\begin{align}
Z_{\rm Nek}^{\rm f}\left(\av,\bfm,\e,\Lambda \right) &= 
Z_{\rm pert}^{\rm f}(\av,\bfm,\e,\Lambda)\sum_{k=1}^\infty 
\sum_{\substack{
 k_1,\cdots,k_N\in \Z_{\ge 0} \\
 k_1+\cdots+k_N=k}}
\sum_{ \vk_1\in Y_{k_1}}
\cdots\sum_{\vk_N\in Y_{k_N}}
Z^{\rm f}_{\rm inst}(\av,\bfm,\bfk,\e,\Lambda), 
\label{fundamental Nekrasov partition}
\end{align}
with
\begin{align}
Z_{\rm pert}^{\rm f}(\av,\e,\Lambda) &= \exp\left\{
\sum_{l \neq n}\gamma_\e(a_l-a_n;\Lambda)
+\sum_{l,f}\gamma_\e(a_l+m_f;\Lambda)
\right\},\\
\label{fundamental inst partition}
Z^{\rm f}_{\rm inst}(\av,\bfm,\bfk,\e,\Lambda)
&=
\Lambda^{k(2N-N_f)} 
\prod_{(l,i)\neq(n,j)}
\frac{a_{ln}+\e(k_{l,i}-k_{n,j}+j-i)}{a_{ln}+\e(j-i)} \nn \\
&\hspace{2.5cm}\times \prod_{l,f,i}
\frac{\Gamma\left(\frac{m_f+a_l}{\e}+k_{li}-i+1\right)}
{\Gamma\left(\frac{m_f+a_l}{\e}-i+1\right)}, 
\end{align}
where $i,j=1,\cdots,\infty$, 
$l,n=1,\cdots,N$, 
$f=1,\cdots,N_f$, 
$a_{ln}\equiv a_l-a_n$, 
$\gamma_\e(x;\Lambda)$ is defined in (\ref{gamma function}), 
and $Y_{l}$ is again a set of Young diagrams with $l$ boxes. 
In \cite{Nekrasov:2003rj}, it was shown that 
(\ref{fundamental inst partition}) 
can be written using the colored profile function 
(\ref{colored profile}) as 
\begin{multline}
 Z_{\rm inst}^{\rm f}\left(\av,\bfm,\bfk,\e,\Lambda \right) =
 \exp\Bigl(
-\frac{1}{4}\int_{x\ne y} dx dy 
f''_{\bfa,\bfk}(x|\e) f''_{\bfa,\bfk}(y|\e) \gamma_\e(x-y;\Lambda) 
\\
+\frac{1}{2}\sum_{f=1}^{N_f} \int dx 
f''_{\bfa,\bfk}(x|\e) \gamma_\e(x+m_f;\Lambda) 
\Bigr). 
\label{fund Nek profile}
\end{multline}

The easiest way to rewrite (\ref{fundamental inst partition}) in 
the Frobenius representation is substituting  
(\ref{extended Frobenius profile}) into the expression 
(\ref{fund Nek profile}). 
The result is
\begin{align}
 Z_{\rm inst}^{\rm f}\left(\av,\bfm,\bfk,\e,\Lambda \right) =
 &Z_{\rm inst}(\bfa,\bfk,\e,\Lambda) \nn \\
 &\times \frac{1}{\Lambda^{kN_f}}
 \prod_{f=1}^{N_f}\prod_{l=1}^N\prod_{i=1}^{L_l}
\Bigl(
{|m_f+a_l-\e\tn_i^l|\cdots|m_f+a_l+\e n_i^l|}
\Bigr),
\label{Nek for fund in Frob}
\end{align}
where $Z_{\rm inst}(\bfa,\bfk,\e,\Lambda)$ is given by 
(\ref{Nekrasov in Frobenius}). 
In deriving (\ref{Nek for fund in Frob}), we have used 
the relation, 
\begin{equation}
 \gamma_\e(x+\e;\Lambda)-\gamma_\e(x;\Lambda) = 
\log\Bigl(\frac{\e^{x/\e}}{\sqrt{2\pi}}
\Gamma(x/\e + 1)
\Bigr). 
\end{equation}

\bibliographystyle{JHEP}
\bibliography{refs}

\providecommand{\href}[2]{#2}\begingroup\raggedright\begin{thebibliography}{10}

\bibitem{Seiberg:1994rs}
N.~Seiberg and E.~Witten, {\it {Electric - magnetic duality, monopole
  condensation, and confinement in N=2 supersymmetric Yang-Mills theory}},
  {\em Nucl. Phys.} {\bf B426} (1994) 19--52
  [\href{http://arXiv.org/abs/hep-th/9407087}{{\tt hep-th/9407087}}].

\bibitem{Seiberg:1994aj}
N.~Seiberg and E.~Witten, {\it Monopoles, duality and chiral symmetry breaking
  in N=2 supersymmetric QCD},  {\em Nucl. Phys.} {\bf B431} (1994) 484--550
  [\href{http://arXiv.org/abs/hep-th/9408099}{{\tt hep-th/9408099}}].

\bibitem{Nekrasov:2002qd}
N.~A. Nekrasov, {\it {Seiberg-Witten prepotential from instanton counting}},
  {\em Adv. Theor. Math. Phys.} {\bf 7} (2004) 831--864
  [\href{http://arXiv.org/abs/hep-th/0206161}{{\tt hep-th/0206161}}].

\bibitem{Losev:2003py}
A.~S. Losev, A.~Marshakov and N.~A. Nekrasov, {\it Small instantons, little
  strings and free fermions},  \href{http://arXiv.org/abs/hep-th/0302191}{{\tt
  hep-th/0302191}}.

\bibitem{Witten:1992xu}
E.~Witten, {\it Two-dimensional gauge theories revisited},  {\em J. Geom.
  Phys.} {\bf 9} (1992) 303--368
  [\href{http://arXiv.org/abs/hep-th/9204083}{{\tt hep-th/9204083}}].

\bibitem{Cordes:1995fc}
S.~Cordes, G.~W. Moore and S.~Ramgoolam, {\it Lectures on 2-d Yang-Mills
  theory, equivariant cohomology and topological field theories},  {\em Nucl.
  Phys. Proc. Suppl.} {\bf 41} (1995) 184--244
  [\href{http://arXiv.org/abs/hep-th/9411210}{{\tt hep-th/9411210}}].

\bibitem{Moore:1997dj}
G.~W. Moore, N.~Nekrasov and S.~Shatashvili, {\it Integrating over Higgs
  branches},  {\em Commun. Math. Phys.} {\bf 209} (2000) 97--121
  [\href{http://arXiv.org/abs/hep-th/9712241}{{\tt hep-th/9712241}}].

\bibitem{Moore:1998et}
G.~W. Moore, N.~Nekrasov and S.~Shatashvili, {\it {D-particle bound states and
  generalized instantons}},  {\em Commun. Math. Phys.} {\bf 209} (2000) 77--95
  [\href{http://arXiv.org/abs/hep-th/9803265}{{\tt hep-th/9803265}}].

\bibitem{Bruzzo:2002xf}
U.~Bruzzo, F.~Fucito, J.~F. Morales and A.~Tanzini, {\it {Multi-instanton
  calculus and equivariant cohomology}},  {\em JHEP} {\bf 05} (2003) 054
  [\href{http://arXiv.org/abs/hep-th/0211108}{{\tt hep-th/0211108}}].

\bibitem{Nekrasov:2003rj}
N.~Nekrasov and A.~Okounkov, {\it Seiberg-Witten theory and random partitions},
   \href{http://arXiv.org/abs/hep-th/0306238}{{\tt hep-th/0306238}}.

\bibitem{Nakajima:2003pg}
H.~Nakajima and K.~Yoshioka, {\it {Instanton counting on blowup. I}},
  \href{http://arXiv.org/abs/math/0306198}{{\tt math/0306198}}.

\bibitem{Nakajima:2003uh}
H.~Nakajima and K.~Yoshioka, {\it {Lectures on instanton counting}},
  \href{http://arXiv.org/abs/math/0311058}{{\tt math/0311058}}.

\bibitem{Maeda:2004is}
T.~Maeda, T.~Nakatsu, K.~Takasaki and T.~Tamakoshi, {\it {Free fermion and
  Seiberg-Witten differential in random plane partitions}},  {\em Nucl. Phys.}
  {\bf B715} (2005) 275--303 [\href{http://arXiv.org/abs/hep-th/0412329}{{\tt
  hep-th/0412329}}].

\bibitem{Billo:2006jm}
M.~Billo, M.~Frau, F.~Fucito and A.~Lerda, {\it {Instanton calculus in R-R
  background and the topological string}},  {\em JHEP} {\bf 11} (2006) 012
  [\href{http://arXiv.org/abs/hep-th/0606013}{{\tt hep-th/0606013}}].

\bibitem{Blumenhagen:2007bn}
R.~Blumenhagen, M.~Cvetic, R.~Richter and T.~Weigand, {\it {Lifting D-Instanton
  Zero Modes by Recombination and Background Fluxes}},  {\em JHEP} {\bf 10}
  (2007) 098 [\href{http://arXiv.org/abs/0708.0403}{{\tt 0708.0403}}].

\bibitem{Ito:2007hy}
K.~Ito, H.~Nakajima and S.~Sasaki, {\it {Deformation of Super Yang-Mills
  Theories in R-R 3-form Background}},  {\em JHEP} {\bf 07} (2007) 068
  [\href{http://arXiv.org/abs/0705.3532}{{\tt 0705.3532}}].

\bibitem{Billo:2007sw}
M.~Billo {\em et.~al.}, {\it {Instantons in N=2 magnetized D-brane worlds}},
  {\em JHEP} {\bf 10} (2007) 091 [\href{http://arXiv.org/abs/0708.3806}{{\tt
  0708.3806}}].

\bibitem{Sasaki:2007iv}
S.~Sasaki, K.~Ito and H.~Nakajima, {\it {Instantons in Deformed Super
  Yang-Mills Theories}},  \href{http://arXiv.org/abs/0710.2218}{{\tt
  0710.2218}}.

\bibitem{Billo':2008sp}
M.~Billo' {\em et.~al.}, {\it {Flux interactions on D-branes and instantons}},
  \href{http://arXiv.org/abs/0807.1666}{{\tt 0807.1666}}.

\bibitem{Billo':2008pg}
M.~Billo' {\em et.~al.}, {\it {Non-perturbative effective interactions from
  fluxes}},  \href{http://arXiv.org/abs/0807.4098}{{\tt 0807.4098}}.

\bibitem{Iqbal:2003ix}
A.~Iqbal and A.-K. Kashani-Poor, {\it {Instanton counting and Chern-Simons
  theory}},  {\em Adv. Theor. Math. Phys.} {\bf 7} (2004) 457--497
  [\href{http://arXiv.org/abs/hep-th/0212279}{{\tt hep-th/0212279}}].

\bibitem{Iqbal:2003zz}
A.~Iqbal and A.-K. Kashani-Poor, {\it {SU(N) geometries and topological string
  amplitudes}},  {\em Adv. Theor. Math. Phys.} {\bf 10} (2006) 1--32
  [\href{http://arXiv.org/abs/hep-th/0306032}{{\tt hep-th/0306032}}].

\bibitem{Eguchi:2003sj}
T.~Eguchi and H.~Kanno, {\it {Topological strings and Nekrasov's formulas}},
  {\em JHEP} {\bf 12} (2003) 006
  [\href{http://arXiv.org/abs/hep-th/0310235}{{\tt hep-th/0310235}}].

\bibitem{Eguchi:2003it}
T.~Eguchi and H.~Kanno, {\it {Geometric transitions, Chern-Simons gauge theory
  and Veneziano type amplitudes}},  {\em Phys. Lett.} {\bf B585} (2004)
  163--172 [\href{http://arXiv.org/abs/hep-th/0312234}{{\tt hep-th/0312234}}].

\bibitem{Hollowood:2003cv}
T.~J. Hollowood, A.~Iqbal and C.~Vafa, {\it {Matrix Models, Geometric
  Engineering and Elliptic Genera}},  {\em JHEP} {\bf 03} (2008) 069
  [\href{http://arXiv.org/abs/hep-th/0310272}{{\tt hep-th/0310272}}].

\bibitem{Zhou:2003zp}
J.~Zhou, {\it {Curve counting and instanton counting}},
  \href{http://arXiv.org/abs/math/0311237}{{\tt math/0311237}}.

\bibitem{Aganagic:2002qg}
M.~Aganagic, M.~Marino and C.~Vafa, {\it {All loop topological string
  amplitudes from Chern-Simons theory}},  {\em Commun. Math. Phys.} {\bf 247}
  (2004) 467--512 [\href{http://arXiv.org/abs/hep-th/0206164}{{\tt
  hep-th/0206164}}].

\bibitem{Iqbal:2002we}
A.~Iqbal, {\it {All genus topological string amplitudes and 5-brane webs as
  Feynman diagrams}},  \href{http://arXiv.org/abs/hep-th/0207114}{{\tt
  hep-th/0207114}}.

\bibitem{Aganagic:2003db}
M.~Aganagic, A.~Klemm, M.~Marino and C.~Vafa, {\it {The topological vertex}},
  {\em Commun. Math. Phys.} {\bf 254} (2005) 425--478
  [\href{http://arXiv.org/abs/hep-th/0305132}{{\tt hep-th/0305132}}].

\bibitem{Okounkov:2003sp}
A.~Okounkov, N.~Reshetikhin and C.~Vafa, {\it {Quantum Calabi-Yau and classical
  crystals}},  \href{http://arXiv.org/abs/hep-th/0309208}{{\tt
  hep-th/0309208}}.

\bibitem{Iqbal:2003ds}
A.~Iqbal, N.~Nekrasov, A.~Okounkov and C.~Vafa, {\it Quantum foam and
  topological strings},  \href{http://arXiv.org/abs/hep-th/0312022}{{\tt
  hep-th/0312022}}.

\bibitem{Tachikawa:2004ur}
Y.~Tachikawa, {\it {Five-dimensional Chern-Simons terms and Nekrasov's
  instanton counting}},  {\em JHEP} {\bf 02} (2004) 050
  [\href{http://arXiv.org/abs/hep-th/0401184}{{\tt hep-th/0401184}}].

\bibitem{Awata:2005fa}
H.~Awata and H.~Kanno, {\it {Instanton counting, Macdonald functions and the
  moduli space of D-branes}},  {\em JHEP} {\bf 05} (2005) 039
  [\href{http://arXiv.org/abs/hep-th/0502061}{{\tt hep-th/0502061}}].

\bibitem{Matsuura:2005sg}
S.~Matsuura and K.~Ohta, {\it {Localization on the D-brane, two-dimensional
  gauge theory and matrix models}},  {\em Phys. Rev.} {\bf D73} (2006) 046006
  [\href{http://arXiv.org/abs/hep-th/0504176}{{\tt hep-th/0504176}}].

\bibitem{Maeda:2005qg}
T.~Maeda, T.~Nakatsu, Y.~Noma and T.~Tamakoshi, {\it {Gravitational quantum
  foam and supersymmetric gauge theories}},  {\em Nucl. Phys.} {\bf B735}
  (2006) 96--126 [\href{http://arXiv.org/abs/hep-th/0505083}{{\tt
  hep-th/0505083}}].

\bibitem{Marshakov:2006ii}
A.~Marshakov and N.~Nekrasov, {\it {Extended Seiberg-Witten theory and
  integrable hierarchy}},  {\em JHEP} {\bf 01} (2007) 104
  [\href{http://arXiv.org/abs/hep-th/0612019}{{\tt hep-th/0612019}}].

\bibitem{Maeda:2006we}
T.~Maeda and T.~Nakatsu, {\it {Amoebas and instantons}},  {\em Int. J. Mod.
  Phys.} {\bf A22} (2007) 937--984
  [\href{http://arXiv.org/abs/hep-th/0601233}{{\tt hep-th/0601233}}].

\bibitem{Iqbal:2007ii}
A.~Iqbal, C.~Kozcaz and C.~Vafa, {\it {The refined topological vertex}},
  \href{http://arXiv.org/abs/hep-th/0701156}{{\tt hep-th/0701156}}.

\bibitem{Marshakov:2007rh}
A.~Marshakov, {\it {On Microscopic Origin of Integrability in Seiberg-Witten
  Theory}},  {\em Theor. Math. Phys.} {\bf 154} (2008) 362--384
  [\href{http://arXiv.org/abs/0706.2857}{{\tt 0706.2857}}].

\bibitem{Taki:2007dh}
M.~Taki, {\it {Refined Topological Vertex and Instanton Counting}},  {\em JHEP}
  {\bf 03} (2008) 048 [\href{http://arXiv.org/abs/0710.1776}{{\tt 0710.1776}}].

\bibitem{Nakatsu:2007dk}
T.~Nakatsu and K.~Takasaki, {\it {Melting Crystal, Quantum Torus and Toda
  Hierarchy}},  \href{http://arXiv.org/abs/0710.5339}{{\tt 0710.5339}}.

\bibitem{Marshakov:2007ra}
A.~Marshakov, {\it {Seiberg-Witten Theory and Extended Toda Hierarchy}},  {\em
  JHEP} {\bf 03} (2008) 055 [\href{http://arXiv.org/abs/0712.2802}{{\tt
  0712.2802}}].

\bibitem{Awata:2008ed}
H.~Awata and H.~Kanno, {\it {Refined BPS state counting from Nekrasov's formula
  and Macdonald functions}},  \href{http://arXiv.org/abs/0805.0191}{{\tt
  0805.0191}}.

\bibitem{Nakatsu:2008mr}
T.~Nakatsu, Y.~Noma and K.~Takasaki, {\it {Integrable Structure of $5d$
  $\mathcal{N}=1$ Supersymmetric Yang-Mills and Melting Crystal}},  {\em Int.
  J. Mod. Phys.} {\bf A23} (2008) 2332--2342
  [\href{http://arXiv.org/abs/0806.3675}{{\tt 0806.3675}}].

\bibitem{Nakatsu:2008px}
T.~Nakatsu, Y.~Noma and K.~Takasaki, {\it {Extended $5d$ Seiberg-Witten Theory
  and Melting Crystal}},  \href{http://arXiv.org/abs/0807.0746}{{\tt
  0807.0746}}.

\bibitem{deHaro:2004id}
S.~de~Haro and M.~Tierz, {\it Brownian motion, Chern-Simons theory, and 2d
  Yang-Mills},  \href{http://arXiv.org/abs/hep-th/0406093}{{\tt
  hep-th/0406093}}.

\bibitem{Matsuo:2004cq}
T.~Matsuo, S.~Matsuura and K.~Ohta, {\it {Large N limit of 2D Yang-Mills theory
  and instanton counting}},  {\em JHEP} {\bf 03} (2005) 027
  [\href{http://arXiv.org/abs/hep-th/0406191}{{\tt hep-th/0406191}}].

\bibitem{Tai:2007vc}
T.-S. Tai, {\it {Instanton Counting and Matrix Model}},  {\em Prog. Theor.
  Phys.} {\bf 119} (2008) 165 [\href{http://arXiv.org/abs/0709.0432}{{\tt
  0709.0432}}].

\bibitem{Hanany:1997ie}
A.~Hanany and E.~Witten, {\it Type IIB superstrings, BPS monopoles, and
  three-dimensional gauge dynamics},  {\em Nucl. Phys.} {\bf B492} (1997)
  152--190 [\href{http://arXiv.org/abs/hep-th/9611230}{{\tt hep-th/9611230}}].

\bibitem{Witten:1997sc}
E.~Witten, {\it Solutions of four-dimensional field theories via M-theory},
  {\em Nucl. Phys.} {\bf B500} (1997) 3--42
  [\href{http://arXiv.org/abs/hep-th/9703166}{{\tt hep-th/9703166}}].

\bibitem{Myers:1999ps}
R.~C. Myers, {\it {Dielectric-branes}},  {\em JHEP} {\bf 12} (1999) 022
  [\href{http://arXiv.org/abs/hep-th/9910053}{{\tt hep-th/9910053}}].

\bibitem{Strominger:1995ac}
A.~Strominger, {\it {Open p-branes}},  {\em Phys. Lett.} {\bf B383} (1996)
  44--47 [\href{http://arXiv.org/abs/hep-th/9512059}{{\tt hep-th/9512059}}].

\bibitem{Giveon:1998sr}
A.~Giveon and D.~Kutasov, {\it Brane dynamics and gauge theory},  {\em Rev.
  Mod. Phys.} {\bf 71} (1999) 983--1084
  [\href{http://arXiv.org/abs/hep-th/9802067}{{\tt hep-th/9802067}}].

\bibitem{Berkooz:1996is}
M.~Berkooz and M.~R. Douglas, {\it {Five-branes in M(atrix) theory}},  {\em
  Phys. Lett.} {\bf B395} (1997) 196--202
  [\href{http://arXiv.org/abs/hep-th/9610236}{{\tt hep-th/9610236}}].

\bibitem{VanRaamsdonk:2001cg}
M.~Van~Raamsdonk, {\it {Open dielectric branes}},  {\em JHEP} {\bf 02} (2002)
  001 [\href{http://arXiv.org/abs/hep-th/0112081}{{\tt hep-th/0112081}}].

\bibitem{Dorey:2002ik}
N.~Dorey, T.~J. Hollowood, V.~V. Khoze and M.~P. Mattis, {\it {The calculus of
  many instantons}},  {\em Phys. Rept.} {\bf 371} (2002) 231--459
  [\href{http://arXiv.org/abs/hep-th/0206063}{{\tt hep-th/0206063}}].

\bibitem{Seiberg:1999vs}
N.~Seiberg and E.~Witten, {\it {String theory and noncommutative geometry}},
  {\em JHEP} {\bf 09} (1999) 032
  [\href{http://arXiv.org/abs/hep-th/9908142}{{\tt hep-th/9908142}}].

\bibitem{Braden:1999zp}
H.~W. Braden and N.~A. Nekrasov, {\it {Space-time foam from non-commutative
  instantons}},  {\em Commun. Math. Phys.} {\bf 249} (2004) 431--448
  [\href{http://arXiv.org/abs/hep-th/9912019}{{\tt hep-th/9912019}}].

\bibitem{Nekrasov:1998ss}
N.~Nekrasov and A.~Schwarz, {\it Instantons on noncommutative R**4 and (2,0)
  superconformal six dimensional theory},  {\em Commun. Math. Phys.} {\bf 198}
  (1998) 689--703 [\href{http://arXiv.org/abs/hep-th/9802068}{{\tt
  hep-th/9802068}}].

\bibitem{LM}
K.~J. Larsen and S.~Matsuura, {\it in preparation}, .

\bibitem{Dijkgraaf:2002fc}
R.~Dijkgraaf and C.~Vafa, {\it Matrix models, topological strings, and
  supersymmetric gauge theories},  {\em Nucl. Phys.} {\bf B644} (2002) 3--20
  [\href{http://arXiv.org/abs/hep-th/0206255}{{\tt hep-th/0206255}}].

\bibitem{Dijkgraaf:2002vw}
R.~Dijkgraaf and C.~Vafa, {\it On geometry and matrix models},  {\em Nucl.
  Phys.} {\bf B644} (2002) 21--39
  [\href{http://arXiv.org/abs/hep-th/0207106}{{\tt hep-th/0207106}}].

\bibitem{Dijkgraaf:2002dh}
R.~Dijkgraaf and C.~Vafa, {\it A perturbative window into non-perturbative
  physics},  \href{http://arXiv.org/abs/hep-th/0208048}{{\tt hep-th/0208048}}.

\bibitem{Katz:1996fh}
S.~H. Katz, A.~Klemm and C.~Vafa, {\it {Geometric engineering of quantum field
  theories}},  {\em Nucl. Phys.} {\bf B497} (1997) 173--195
  [\href{http://arXiv.org/abs/hep-th/9609239}{{\tt hep-th/9609239}}].

\bibitem{Katz:1997eq}
S.~Katz, P.~Mayr and C.~Vafa, {\it {Mirror symmetry and exact solution of 4D N
  = 2 gauge theories. I}},  {\em Adv. Theor. Math. Phys.} {\bf 1} (1998)
  53--114 [\href{http://arXiv.org/abs/hep-th/9706110}{{\tt hep-th/9706110}}].

\end{thebibliography}\endgroup

\end{document}